\documentstyle[psfig,conf_iap,]{article}
\def\HI{H~{\sc i}}
\def\HII{H~{\sc ii}}

\def\HeI{He~{\sc i}}
\def\NI{N~{\sc i}}
\def\OI{O~{\sc i}}
\def\NeI{Ne~{\sc i}}
\def\MgI{Mg~{\sc i}}
\def\MgII{Mg~{\sc ii}}
\def\CII{C~{\sc ii}}

\def\CIIstar{C~{\sc ii}$^*$}
\def\CaII{Ca~{\sc ii}}
\def\deeg{$^{\rm o}$}
\def\kms{km s$^{-1}$}
\def\cmtwo{cm$^{-2}$}
\def\cc{cm$^{-3}$}
\def\nel{n(e)}
\def\nHI{n(H~{\sc i})}
\def\nHeI{n(He~{\sc i})}

\begin{document}

\heading{%
%Begin Heading
%
%%
Local Interstellar Matter: The view from Paris\\
%
%End Heading
} 
\par\medskip\noindent
\author{%
%Begin Author names
Priscilla C. Frisch$^{1}$
%End Author names
}
\address{%
Department of Astronomy and Astrophysics, University of Chicago,
5640 South Ellis Avenue, Chicago, IL  60637
}

\begin{abstract}
Observations of interstellar gas and dust towards nearby stars and
within the solar system show that the Sun is embedded in a 
warm diffuse partially-ionized cloud.  This cloud is the leading edge
of a flow of interstellar matter (ISM) through our Galactic neighborhood.
Interstellar matter sets the boundary conditions of the heliosphere
and astrospheres of extra-solar planetary systems.  Moreover,
the interplanetary regions in these systems are sensitive to inflowing ISM.
\end{abstract}
\section{Introduction} 

What better place to review the Local Interstellar Medium (LISM) than
in Paris -- which has nurtured creative thinking in the
sciences, mathematics and arts for many centuries?

The launch of the $Copernicus$ satellite in 1972 provided UV spectrometers capable
of observing dominant trace ions in the low density intercloud
medium (as it was then called) around the Sun. 
York, Jenkins, Spitzer and colleagues
discovered that gas towards $\alpha$ Leo (24 pc) is $\sim$50\% ionized
since N(N~{\sc i})$\sim$N(N~{\sc ii}) \cite{RogersonIII:1973}, while
observations of nearby cool stars demonstrated the low
mean densities of the closest ISM (n$\leq$0.1 \cc, e.g.
\cite{McClintocketal:1976}).
$Copernicus$ yielded the first spectral observations of
interstellar \HI\ inside of the solar system, showing that
this material has a velocity several \kms\ different from
the velocities of nearby interstellar clouds located in the upwind direction \cite{AdamsFrisch:1977}.
Vidal-Madjar and colleagues
used $Copernicus$ data to evaluate the inhomogeneity
of ISM within a few parsecs, based on a possible gradient in radiation pressure 
\cite{VidalMadjaretal:1978}.
The fact that the interstellar cloud seen inside of the solar
system (now called the Local Interstellar Cloud, LIC) does $not$ extend to 
$\alpha$ Cen, the nearest star, was originally discovered with $Copernicus$
\cite{Landsmanetal:1984} and remains a puzzle.
During the last decade of the 20th century, observations by GHRS and 
STIS on HST, and high-resolution optical
data, provided the basic properties, composition, and kinematics of individual
cloudlets close to the Sun.  $Ulysses$ and $Galileo$ have now made
{\it in situ} observations of interstellar gas and dust inside of the solar system,
and the pickup ions and anomalous cosmic ray populations
which result from the interaction of ISM with the solar wind.

These data, together with heliosphere\footnote{The heliosphere is the region of space containing the solar wind plasma.} models,
show that the physical properties of the ISM influence
the heliosphere configuration, the astrospheres of external cool
stars, and the interplanetary environment of both our own solar system
and extra-solar planetary systems \cite{Frisch:1993a,ZankFrisch:1999}.

\section{LISM Gas towards Nearby Stars}

The relative absence of O/B stars nearby the Sun (there are no O stars and four B-stars within 30 pc) means that
the LISM properties are pieced together from
observations of several types of target objects, each
with limitations, including 
A, F, G, K stars (there are 57 A stars within 30 pc)
and white dwarf stars (118 are within 20 pc \cite{HolbergOswaltSion:2001}).

Optical \CaII\ and UV absorption lines
show that the Sun is embedded in a flow of interstellar cloudlets with 
a bulk velocity of --28.2 \kms, and approaching the Sun
from the upstream direction l=12.6\deeg, b=11.7 \deeg\
\cite{FrischWeltyGrodnicki:2001}.  This corresponds to an upwind
direction in the local standard of rest of l=3\deeg, b=--5\deeg, and flow velocity --17.0 \kms.\footnote{Using the Hipparcos 
value for the solar apex motion \cite{DehnenBinney:1998}.
} 
The distribution of absorption components about the best-fit velocity
is consistent with a Gaussian (e$^{-(V - V_o)^2/2 \sigma ^4}$) 
with a dispersion $\sigma  \sim$4.5 \kms.
The strongest \CaII\ absorption lines within 30 pc are found
towards $\alpha$ Oph (14 pc), but the properties of the cloud forming
these features are virtually known.
Three \CaII\ components are seen towards $\alpha$ Aql (5 pc), showing the complex
LISM structure \cite{Ferlet:1986}.  
Based on cloud radial velocities (V), the next interstellar cloud
to be encountered by the Sun as it traverses interstellar space is likely to be either the cloud
in front of $\alpha$ Cen (V$\sim$--18 \kms) or one of the clouds 
towards $\alpha$ Aql (e.g. V$\sim$--27 \kms) located
near the direction of solar apex motion. 

The ionization of the LISM has been found from
observations of \MgI/\MgII\ and \CIIstar/\CII\ 
towards nearby stars.
The \CIIstar\ fine-structure lines are insensitive to the
radiation field since they are collisionally populated,
while in warm gas \MgI\ is dominated by photoionization, and 
radiative and dielectronic recombination.
Values range from \nel=0.12 \cc\ (\CIIstar\ data) to
\nel=0.35 \cc\ (\MgI\ data),
based on observations of
$\alpha$ CMa \cite{Lallementetal:1994}, 
$\epsilon$ CMa \cite{GryJenkins:2001},
REJ 1032+532 \cite{Holberg:1999} 
$\delta$ Cas \cite{LallementFerlet:1997}, 
$\eta$ UMa \cite{Frischetal:2001},
and including unpublished values for
two components  towards $\alpha$ Aql
(\nel=0.35$\pm$0.16 \cc\ and 0.15$\pm$0.10 \cc).  
The difference between \CIIstar\ and \MgI\ results may indicate that \MgI\ and \MgII\
do not sample identical portions of the clouds.

\begin{figure}
\centerline{\vbox{
\psfig{figure=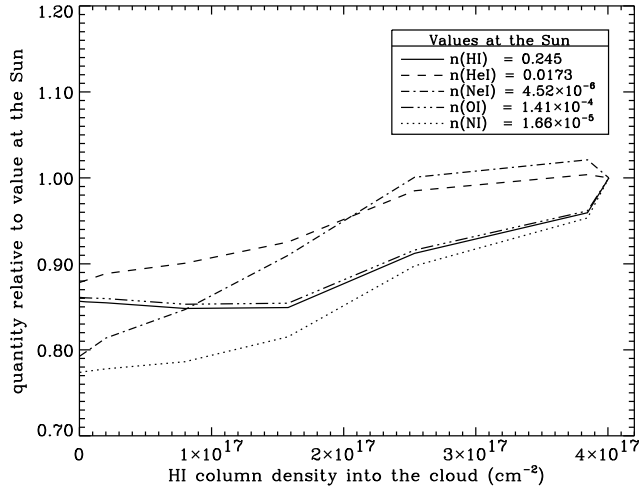,height=7.cm}
}}
\caption[]{
Predictions for the densities of H~{\sc I}, He~{\sc I}, Ne~{\sc I}, O~{\sc I} N~{\sc I},
relative to those at the Sun as a function of N(H~{\sc I}) depth
into the LIC.
The cloud surface is at the left and the solar location is at the
right (from SF17, see text).  \label{slavinfrisch}  
}
\end{figure}

A self-consistent photoionization model of interstellar matter 
within $\sim$ 5 pc of the Sun yields \nel$\sim$0.13 \cc, \nHI$\sim$0.24 \cc, 
T$\sim$7200 K, and fractional ionization X(H)$\sim$0.3 and X(He)$\sim$0.4
at the solar location, in good agreement with observations
(Model 17 in \cite{SlavinFrisch:2001}, hereafter SF17).
This model, constrained both by {\it in situ} particle data and absorption line data, 
includes radiation from nearby hot stars, diffuse emission from the soft 
X-ray background, and unobserved EUV emission from a possible evaporative 
boundary between the local ISM and hot gas (also see Slavin's talk in
this volume).  
In low column-density gas \NI\ equilibrium depends partly on
photoionization, while at higher column
densities N and H ionization are coupled by charge exchange.
In contrast, O and H ionization are tightly coupled by charge
exchange.  
At lower column densities, SF17 predict an inverse relation between N(O~{\sc I})/N(N~{\sc I}) and
N(H~{\sc I}) (Figure \ref{slavinfrisch}), and this same behavior is
observed for Log N(H~{\sc I})$>$17.8 \cmtwo\ (Figure \ref{NO}).  
%The observed inverse relation between N(O~{\sc I})/N(N~{\sc I}) and
%N(H~{\sc I}) seen for Log N(H~{\sc I})$>$17.8 \cmtwo\ (Figure \ref{NO}) 

The LISM shows enhanced abundances for
refractory elements typical of warm disk gas which has been
processed by shock front activity \cite{Frischetal:1999}.
The photoionization model (above) compares abundances to 
the total of \HI\ and \HII.  Resulting abundances for C, N, O within 5 pc are,
respectively, 427, 81, and 630 parts-per-million by number (SF17).

Observations towards local white dwarfs (d$<$100 pc) at wavelengths $<$912 \AA\ show that the interstellar \HI/\HeI\
ratio varies by a factor of $\sim$2 between sightlines
(e.g.  \cite{Vallerga:1996}), indicating the ionization of LISM is
inhomogeneous. 

ISM within 30 pc is distributed asymmetrically around the Sun, 
and the Sun is located in the leading edge of an outflow from
Loop I and the Sco-Cen Association \cite{Frisch:1995}.

\section{Observations of LISM Gas inside of the Heliosphere}

ISM inside of the heliosphere provides a unique window on the LIC.
Interstellar neutrals flow into the
heliosphere and are ionized near $\sim$5 AU for \HI\ (mainly by charge
exchange with the solar wind) and $\sim$0.5 AU for \HeI\ (by photoionization).
The detection by $Ulysses$ of interstellar \HeI\ provides a direct
measure of the heliocentric velocity of the LIC \cite{Witte:1996}
which agrees with values found from
the FUV fluorescent scattering of solar 584 \AA\ emission \cite{Flynne:1998}.
The LIC \HeI\ density, temperature and heliocentric velocity are 
\nHeI=0.017 \cc, T=6700$\pm$900, and v=--25.5 \kms,
and this ``local interstellar wind'' arrives from the upstream
direction of l=2.7 \deeg, b=15.6 \deeg.
Absorption components near the LIC velocity are identified
towards 30 nearby stars, but surprisingly $not$ towards the nearest
star $\alpha$ Cen. 

\begin{center}
\begin{figure}
\centerline{\vbox{
\psfig{figure=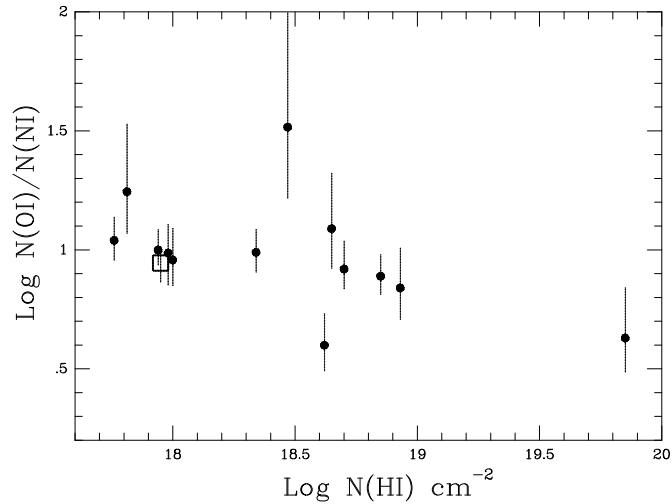,height=4.5cm}
}}
%\vspace{-0.5in}
\caption[]{Log N(O~{\sc I})/N(N~{\sc I}) versus Log N(H~{\sc I})
for stars sampling nearby ISM.  Data points are for
G191-B2B, HZ 43, WD 0621-376, WD 2211-495,
BD +28D 4211,WD 1634-573), Feige 24, GD394, WD 2331-475, Sirius, REJ 1032+532,
$\epsilon$ CMa, and $\eta$ UMa, based on FUSE, HST (GHRS and STIS), and IUE data
\cite{Holberg:1999,Hebrardetal:1999,Vennesetal:2000,Friedmanetal:2001,Lemoineetal:2001,Lehneretal:2001,Kruketal:2001,Hebrardetal:2001,Sonnebornetal:2001,Woodetal:2001,JenkinsetalAr:2000,GryJenkins:2001,Frischetal:2001}.  
The points at (x,y)=(18.6,0.6), (17.8,1.2), and (18.5,1.5) are, respectively, REJ 1032+532,
Sirius, and Feige 24.
The ratio N/O for pickup ions (the box) is plotted at Log N(H~{\sc I})=17.8 \cmtwo\
(SF17).
\label{NO}}
\end{figure}
\end{center}

Neutral ISM flows into the heliosphere, interacts with the solar
wind plasma, and forms the pickup ion (PUI) 
and anomalous cosmic ray populations.
The PUIs give the composition of the parent population of interstellar \OI, \NI, \NeI, and \HeI\ atoms 
\cite{GloecklerGeiss:2001}, and seed
the anomalous cosmic rays population observed in the inner solar
system (e.g. \cite{CummingsStone:1996}).  The PUIs
yield O~{\sc I}/N~{\sc I}=7.0$\pm$1.5 inside of the heliosheath, or
8.8$\pm$1.5 after correcting for 25\% heliosphere filtration of \OI.
Abundance ratios in both PUIs and the anomalous cosmic-rays
can be used to constrain LIC ionization \cite{Frisch:1994,SlavinFrisch:2001}, 
and observations of nearby stars show that results from
observations inside and outside of the heliosphere are consistent
(Figure \ref{NO}).

\begin{center}
\begin{figure}[h]
\centerline{\vbox{
\psfig{figure=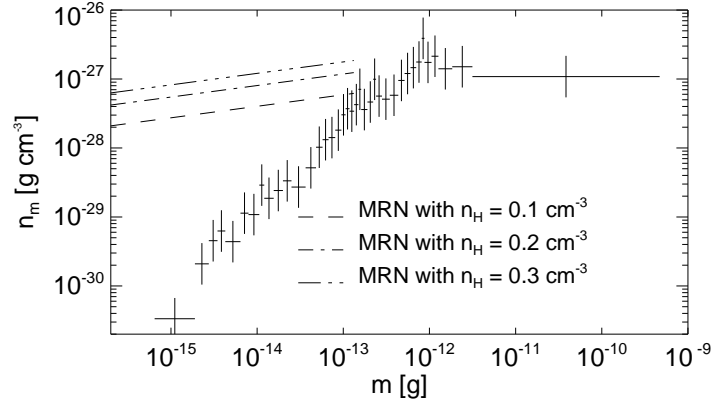,height=6.cm}
}}
\vspace{-0.15in}
\caption[]{The observed mass-density distribution for
interstellar dust grains observed within the solar system
(\cite{Frischetal:1999}).  The dashed lines show the MRN distribution
\cite{Mathis:1977} for three normalizations of the total mass density in the LIC. 
The vertical line gives the
upper limit cutoff of the MRN distribution  
(see \cite{Frischetal:1999} for additional information).
\label{dust}
}
\end{figure}
\end{center}

$Ulysses$ and $Galileo$ have observed interstellar dust grains
with masses 10$^{-15}$ to 10$^{-10}$ g,
flowing through the solar system at the LIC velocity (Figure \ref{dust},
\cite{Frischetal:1999}).
The gas-to-dust mass ratio for the measured grain population is
consistent with values for the general ISM.
For n$_{\rm total}$=0.37 \cc\ (SF17), the observed grains correspond 
to a gas-to-dust mass ratio of 116$^{+46}_{-38}$.  However $\sim$40\% of the 
grain mass is removed by
%ratio of 94$^{+46}_{-38}$.  However $\sim$40\% of the grain mass is removed by
heliosheath\footnote{The heliopause is the contact discontinuity between the solar wind and
interstellar plasmas.  The deflected
plasmas on either side of the heliopause form the inner and outer
heliosheath regions.} 
and heliospheric filtration of small charged grains interacting with the
solar wind, indicating a lower value (R$_{\rm g/d}$$\sim$70) for the LIC (assuming H/He=10).

\section{Conclusions}

ISM sets the boundary conditions of the heliosphere and the
astrospheres of extra-solar planetary systems.
In the low density environment of the Sun today,
interstellar gas surprisingly comprises $\sim$98\% of the diffuse material within
the heliosphere.
An increase in the LIC density from the current value 
to n(\HI)=10 \cc\ would cause the heliosphere to contract to a radius of $\sim$10--15 AU
 in the
upstream direction (from the current value of $\sim$100 AU), dramatically alter 
the interplanetary environments of the inner planets,
and expose outer planets to ``raw'' ISM  \cite{ZankFrisch:1999}.  Although the
consequences of modifying the heliosphere configuration are unknown
(despite ``black cloud'' scenarios),
speculations about climate modifications from such an encounter have long
intrigued scientists \cite{Shapley:1921} and challenged our 
understanding of the Galactic environment of the Sun.

\vspace{0.5in}

\vspace{-0.8in}
\acknowledgements{The author would like to thank NASA for 
support through grants NAG5-6405 and NAG5-7077.}

\vfill
\end{document}